\documentclass[runningheads]{llncs}
\usepackage[T1]{fontenc}
\usepackage{graphicx}
\usepackage{cite}
\usepackage{comment}
\usepackage{amsmath,amssymb,amsfonts}
\usepackage{algorithmic}
\usepackage{textcomp}
\usepackage{xcolor}
\usepackage{subcaption}
\usepackage{listings}
\usepackage{url}
\usepackage{hyperref}
\usepackage{ulem}
\usepackage{todonotes}
\usepackage{xcolor}

\newcommand{\orcid}[1]{\href{https://orcid.org/#1}{\includegraphics[height=10pt]{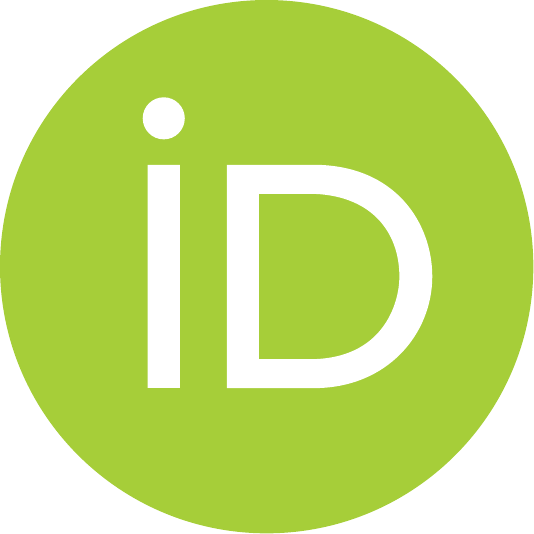}}}

\def\BibTeX{{\rm B\kern-.05em{\sc i\kern-.025em b}\kern-.08em
    T\kern-.1667em\lower.7ex\hbox{E}\kern-.125emX}}

\begin{document}
\title{A New Execution Model and Executor for Adaptively Optimizing the Performance of Parallel Algorithms Using HPX Runtime System}

\author{Karame Mohammadiporshokooh \inst{1,2}\orcidID{0009-0000-8349-3389}\and
Steven R. Brandt \inst{1,2} \orcidID{0000-0002-7979-2906} \and
Hartmut Kaiser \inst{1,2}\orcidID{0000-0002-8712-2806}}
\authorrunning{K. Mohammadiporshokooh et al.}

\institute {Center of Computation and Technology, Louisiana State University, Baton Rouge, LA, 70803, USA 
\and Department of Computer Science, Louisiana State University, baton Rouge, LA, 70803, USA\\
Corresponding author(s)\email : {kmoham6@lsu.edu} \\
Contributing authors(s) \email :{\{sbrandt,hkaiser\}@cct.lsu.edu}}
\maketitle

\begin{abstract}
 Developing parallel algorithms efficiently requires careful management of concurrency across diverse hardware architectures. C\texttt{++} executors provide a standardized interface that simplifies the development process, allowing developers to write portable and uniform code. However, in some cases, they may not fully leverage hardware capabilities or optimally allocate resources for specific workloads, leading to potential performance inefficiencies. Building on our earlier conference paper [Adaptively Optimizing the Performance of HPX's Parallel Algorithms], which introduced a preliminary strategy based on cores and chunking (workload), and integrated it into HPX’s executor API, that dynamically optimizes for workload distribution and resource allocation, based on runtime metrics and overheads, this paper, introduces a more detailed model of that strategy. It evaluates the efficiency of this implementation (as an HPX executor) across a wide range of compute-bound and memory-bound workloads on different architectures and with different algorithms. The results show consistent speedups across all tests, configurations, and workloads studied, offering improved performance through a familiar and user-friendly C\texttt{++} executor API. Additionally, the paper highlights how runtime-driven executor adaptations can simplify performance optimization without increasing the complexity of algorithm development

\keywords{Asynchronous Many-Task (AMT) \and HPX \and Executors \and Performance \and Optimization \and Parallel Algorithms} 
\end{abstract}

\section{Introduction}
Managing parallelism and concurrency has become increasingly complex in
software due to the heterogeneity of modern
hardware.  HPX provides executors to offer uniform APIs for execution of
tasks across different architectures. While this
abstraction eases development, it delegates many runtime metrics, such as number of cores or chunk sizes, to the execution policy, where poor choices can lead to performance degradation. For instance, even a straightforward embarrassingly parallel operation on a multi-core processor can be sensitive to the resource allocation strategy. While distributing work across all cores may be effective for larger input data, utilizing more cores for smaller input sizes may lead to performance degradation.

Beyond hardware considerations, there is a
wide range of parallel algorithm types, including map-type algorithms (e.g.
\texttt{copy}, \texttt{fill}, stencil updates, etc.), map-reduce-type algorithms
(e.g. \texttt{min\_element}, \texttt{all\_of}, \texttt{count}, and prefix sums,
etc.), to name a few. The performance of these algorithms can vary depending on
several factors, such as the nature of the algorithm (whether it is
compute-bound or memory-bound), the number of iterations, the number of cores and threads, data
partitioning, and the specific architecture (cache sizes, memory bandwidth, etc.).
The challenge is that no single set of parameters can guarantee
optimal performance across all scenarios. The optimal configuration for parallel
execution is often highly dependent on the specific workload and runtime
environment.

Ideally, parallel speedup should scale linearly with the number of processors~\cite{ESMP2}. However, as Amdahl's Law~\cite{amdahl1967validity} has shown, the maximum speedup of a process on multiple processors is limited by its sequential portion. This limitation can be mitigated by increasing the amount of parallel work as more processors are available, as described by Gustafson's Law~\cite{gustafson1988reevaluating}. More generally, in the context of parallelization, there are four factors to consider:
\begin{itemize}
    \item \textbf{Starvation:} Occurs when there's not enough concurrent work available, leading to reduced resource utilization.
    \item \textbf{Latencies:} Refers to time-distance delays in accessing remote resources. 
    \item \textbf{Overheads:} Refers the work required for managing parallel operation, a feature absent in sequential codes.
    \item \textbf{Waiting for contention resolution:} Delays due to the unavailability of shared resources.
\end{itemize}

 All these factors contribute to reduced parallel efficiency. This paper focuses on reducing overheads. We study the effects on performance of workload distribution (chunk size) and the number of processors utilized.

This paper addresses these challenges by proposing a dynamic optimization
approach integrated into HPX's executor API.
We propose an aggregated model that combines effects such as architectural specifics to get the best possible performance regarding the number of cores and chunk sizes. Our focus is on optimizing workload
distribution and resource allocation through runtime adaptivity.
Unlike approaches that try to break down overhead into many separate factors (cache effects, memory bandwidth, etc.), our model combines all those contributing factors as a single, combined overhead value $T_0$. 
We measure this combined overhead per loop invocation on each target architecture instead of attempting to calculate it.
This method
aims to reduce overheads, such as those associated with parallelism management,
while maintaining the simplicity of user-friendly APIs.
In our previous paper~\cite{10.1007/978-3-031-97196-9_6}, we introduced an adaptive performance model and examined the impact of varying core counts and chunk sizes on two algorithms: adjacent-difference and artificial-work. In this paper, we extend our analysis by running additional benchmarks on diverse architectures to validate the model’s architecture‑independence. We also explore a wider range of chunking configurations to confirm that our identified optimal chunking remains effective. Finally, we evaluate our enhanced approach on an algorithm implemented in ChplX~\cite{atre2025closingsourcecomplexitygap}, which is a source to source translator that converts a Chapel~\cite{1299190} program into a C\texttt{++} program using HPX. 

We use HPX, a C\texttt{++} standard library for parallelism and
concurrency, as the framework for this study, because it is easily customizable and open-source.
Furthermore, it offers similar semantics and performance to OpenMP scheduling for comparable
problems. Indeed, HPX has even been used to build a backend for OpenMP~\cite{zhang2019introduction1}.

Our results demonstrate
that adapting execution configurations based on runtime metrics can prevent the
performance degradation often associated with generic abstractions like
executors or OpenMP's \textit{parallel for} pragma.

\section{Motivation}
\label{sec:motivation}
Scientific simulation codes are often based on Cauchy problems, i.e., codes described by an initial data on an N-dimensional spatial grid, and accompanied by a system of differential equations used to iteratively evolve the data through time. These codes, common in science and industry, typically involve the use of finite differences to simulate spatial derivatives, operations that are very much like the \textit{adjacent-difference} algorithm that calculates the difference between each pair of elements within a sequence. Such time integration methods and algorithms often expose sufficient parallel work that can benefit from parallel computational constructs like parallel loops or parallel regions of independent tasks. On CPUs, parallelizing loops is performed by splitting (chunking) the computational grid such that the parts of the grid can be independently and concurrently scheduled on separate processing units (cores). Determining the size of those chunks and the number of processing units to utilize that result in the best possible execution times is---more often than not---left to the programmer. Selecting the optimal parameters for these parallel loops requires application specific measurements and manual tuning for each of the parallel constructs involved. Common off-the-shelf runtimes like OpenMP and the C\texttt{++}17 Standard library expect the user to tune and provide such metrics. This problem is aggravated when using library solutions like the C\texttt{++}17 Standard algorithms, as the optimal size of the chunks and optimal number of cores depend on the size of the computational space and the amount of work in the user-supplied element access functions passed to those algorithms. Naturally, this amount of work is unknown to the implementer of the parallel algorithms. The optimal values for the parallelization parameters also depends on system- and implementation specific characteristics, e.g. the overhead per scheduled task, the general amount of---possibly nested---parallelism intrinsic to the application, the underlying computing architecture, and many more factors~\cite{7307668}. 

To address this problem, we describe a methodology for adaptively determining the optimal chunk sizes and the number of processing units used for HPX's parallel C\texttt{++} Standard algorithms. We have developed a mathematical model that we use to dynamically find the number of cores and chunk sizes for map-type parallel algorithms that result in the best possible execution performance of that algorithm. We have implemented this model as an HPX execution policy and studied its performance on a variety of algorithms.

In order to achieve this goal,
we measure the time to execute the first invocation of the loop.
The measurement is passed to the model to calculate the best core-chunk configuration.
The result is used by all subsequent invocations
of the same parallel loop.
This approach reduces the measurement overhead required to find the time per loop execution for a particular platform and execution environment by amortizing this overhead over all subsequent calls of the loop.

The required functionality is encapsulated in a simple C\texttt{++} type that serves as the customization point for HPX's parallel algorithms, and so it does not require us to modify the algorithms' implementation. This will be explored in greater depth in section~\ref{subsec:algorithms}.

Some recent studies also address chunking, scheduling, and load balancing for parallel loops. For example, the Distributed Chunk Calculation Approach (DCA) compares centralized vs distributed chunk size calculation under CPU slowdown scenarios, showing gains when chunk sizing is made more adaptive to runtime conditions~\cite{ELELIEMY2021101284}. Likewise, Chunks and Tasks provides abstractions for splitting both data and work into chunks and tasks with dynamic work distribution, which helps with irregular workloads and balancing~\cite{DBLP:journals/corr/abs-1210-7427}.  In a similar work ~\cite{10.1145/3620665.3640405} introduce the Heartbeat Compiler (HBC), which translates C/C++ programs with nested parallelism into binaries capable of automatic granularity control, effectively managing overheads in irregular workloads

While these approaches enhance parallelism through adaptive chunking and scheduling, they do not incorporate runtime measurements to derive overhead and sequential execution times. Our work differs by measuring early loop runtimes to estimate these parameters, enabling dynamic adaptation of both core count and chunk size based on empirical data, thereby optimizing performance in real-time. 

\section{Theory}
\label{sec:theory}




 To start, we define $T_1$ as the total time taken to execute a loop on a single thread without parallelism, and $T_N$ as the total time with $N$ being the number of threads (assuming $N>1$) used to execute the loop. We now assume that the loop can be perfectly parallelized apart from a constant overhead, $T_0$.

\begin{equation}
\label{eq:t_n}
T_N = \frac{T_1}{N} + T_0
\end{equation}

Alternatively, we can derive the relation between speedup (that measures how much faster a parallel algorithm runs compared to a sequential one), efficiency  (that measures how well a system utilizes parallelism), and number of processors to calculate the optimal number of processors from first principles.

We can now compute the speedup as follows:

\begin{equation}
\label{eq:s}
S = \frac{T_1}{T_N}
\end{equation}

Thus far, the speedup formula is generic to any model of computation, including Amdahl's Law and Gustafson's Law. However, substituting Equation~\ref{eq:t_n} into Equation~\ref{eq:s}, we obtain the following:

\begin{equation}
\label{eq:olaw}
S = \frac{T_1}{\frac{T_1}{N} + T_0}
\end{equation}

We point out that this ``Overhead Law'' is different than Amdahl's Law and Gustafson's Law. The former assumes that a fixed fraction of the code is serial, the latter assumes a fixed amount of serial code is always present. In our case, we assume that a fixed amount of serial code is only present when parallelism is attempted. If one wished to create a formula more comparable to these other two laws, one could write the parallel fraction for our case as $p = T_1/(T_0+T_1)$. Substituting this yields:

\begin{equation}
\label{eq:olaw2}
S = \frac{p}{1-p+\frac{p}{N}}
\end{equation}

So the "Overhead Law" differs from Amdahl's Law by a constant factor. We point out that, unlike the other laws, this equation is not valid for the case $N=1$ since Equation~\ref{eq:t_n} only applies when $N>1$.

Now that we have the ``Overhead Law,'' we can compute the efficiency, which is defined as the ratio of speedup to the number of threads: 

\begin{equation}
\label{eq:e1}
E = \frac{S}{N}
\end{equation}

Based on Equation~\ref{eq:s} and~\ref{eq:e1} we can derive the following: 

\begin{equation}
\label{eq:e2}
E = \frac{S}{N} = \frac {T_1}{N T_N}
\end{equation}

After using Equation~\ref{eq:t_n} and substituting for $T_N$, we can simplify and solve for N: 

\begin{equation}
\label{eq:n}
N = \frac{1-E}{E} * \frac {T_1}{T_0}
\end{equation}

We will, somewhat arbitrarily, choose an efficiency (E) of 95\%. A different choice could be made, and the remainder of the analysis would change accordingly.

Note that, using Eq.~\ref{eq:n}, we can determine that $T_{opt} = 19 T_0$ in Eq.~\ref{eq:n_c}. In other words, if we allocate chunks of size $19 T_o$ to each core, we will achieve $95\%$ efficiency. 



Alternatively, we can compute the optimal core count, $N$, using $T_{opt}$.

\begin{equation}
\label{eq:n_c}
N\:=\:{\frac{T_1}{T_{opt}}}
\end{equation}

where as before, $T_1$ is the total time taken per workload and $T_{opt}$ is the time for minimum work per core. 
If we find the amount of work where parallelization with 2 cores gives a speedup of $1.9$ (a $95\%$ efficiency), based on our experiments we can, with the given formula, calculate $T_{opt}$. This will result in $T_{opt}=19 T_0$ and provide the same answer as Equation~\ref{eq:n}.

Once the optimal core count for a given workload is determined, we need to find the optimal chunk size. To find that, 
we use the following formula to calculate the minimum workload for each chunk:

\begin{equation}
\label{eq:t_m}
T_m\:=\:{\frac{T_1}{N * C}}
\end{equation}

Where $C$ is chunks-per-core (which is equal to 8 based on the experiments), $T_1$ is the overall time per workload and $T_m$ is the amount of work per chunk. Assuming that $T_1$ and $T_m$ are both proportional to the number of elements they contain, we can derive an equation for the chunk size ($N_{Chunk}$). The value of $N_E$ is the number of elements in our workload.

\begin{equation}
\label{eq:t_nc}
N_{Chunk}\:=\:{\frac{N_E}{N * C}}
\end{equation}

This equation ensures that $C=8$ chunks per core are used for any workload, with the chunk size always being at least $N_{Chunk}$.

The remaining question is 'How can $T_0$ and $T_1$ be estimated from measurements? To obtain concrete values for $T_0$
 (the overhead) and $T_1$
 (time for sequential work), we measure $T_N$
 for several thread counts 
N>1. We assume the model
\begin {equation}
  T_N = T_0 + T_1/N
 \end {equation}
 and use ordinary least squares regression to find $T_0$ and $T_1$ that best fit the observed data (minimizing the sum of squared differences between predicted and measured runtimes)~\cite{Hofinger2017ParallelOverhead}. Once these parameters are estimated, they are substituted into Equations~\ref{eq:t_n} to Equation~\ref{eq:t_nc} to drive predictions of speedup, efficiency, and optimal configurations in our model. 

Another practical way to measure $T_0$ is to measure the loop once in sequential mode and once in parallel with one core. Then the difference between these two values is the $T_0$ per core which we need for our model. This is identical to the approach we mentioned above.

\section{Implementation}
\label{sec:imp}

Now that we know how to compute the ideal number of cores, $N$, we turn to the implementation. We use HPX because it is available as an open source library, easily customizable, and well-optimized. For this work, we will consider the parallelization of algorithms and loops, scheduling work in a manner similar to the familiar \textit{static} scheduling of OpenMP.

We will start this section with some background on HPX, executors, and customization points.

  \subsection{HPX}

HPX is a C\texttt{++} Standard Library for parallelism and concurrency~\cite{hpx_joss_paper,kaiser_2024_598202}. HPX is an asynchronous many-task runtime (AMT) system that is built on the ParalleX execution model~\cite{paper5364511}. It is implemented as a lightweight user-level task manager running on top of kernel threads~\cite{taskbench}. It is widely known that thread creation and destruction managed by the operating system are expensive and reserve lots of memory~\cite{taskbench}. For that reason, HPX creates and binds one kernel thread (running a user-level queue with work-steeling) to each core~\cite{paper5364511}. User-level threads (i.e. tasks) can be shared among these queues and executed in parallel. HPX thereby abstracts the writing of parallel code from the number of cores available on the machine.

In addition, HPX suspends threads instead of blocking for mutexes, waiting on futures, etc. This means that HPX is able to assign one worker per core and keep the workers busy even for applications that, semantically, are blocking.

The HPX asynchronous programming model exposes a C\texttt{++} standard API entirely conforming to interfaces as defined by the C\texttt{++} Standard and extends the standardized APIs by providing distributed and heterogeneous computing based on \textit{Futures}, which makes HPX portable and uniformly usable for local and remote parallelism~\cite{taskbench}. 

In this paper we focus on HPX's customization points for the control of parallelism within the algorithms to implement our newer, more dynamic method.

\subsubsection{HPX Algorithms and Customization}
\label{subsec:algorithms}

\textbf{Execution policies} are objects that can be provided as the first argument to the standard algorithms of C\texttt{++}. The C\texttt{++}17 and C\texttt{++}20 Standards introduce four execution policy types~\cite{cxx17_standard,cxx20_standard}. All of these are used for unique predefined type instances to disambiguate parallel algorithm overloading:   
\begin{itemize}
    \item \texttt{std::execution::seq} requires that a parallel algorithm's execution not be parallelized. The invocations of element access functions in parallel algorithms are indeterminately sequenced in the calling thread.
    \item \texttt{std::execution::par} indicates that a parallel algorithm's execution may be parallelized. The invocations of element access functions are permitted to execute in either the invoking thread or in a thread implicitly created by the library and are indeterminately sequenced with respect to each other.
    \item \texttt{std::execution::par\_unseq} indicates that a parallel algorithm's execution may be parallelized, vectorized, or migrated across threads. The invocations of element access functions are permitted to execute in an unordered fashion in unspecified threads, and unsequenced with respect to one another within each thread.
    \item \texttt{std::execution::unseq} indicates that a parallel algorithm's execution may be vectorized, e.g., executed on a single thread using instructions that operate on multiple data items.
\end{itemize}

In HPX, we have implemented all of the above with correct semantics in the \texttt{namespace hpx::execution}. However, while implementing all of the parallel algorithms as specified by the C\texttt{++} Standard, the community quickly realized that the execution policies alone do not provide sufficient flexibility for controlling the execution environment. Early discussions in the ISO standardization committee documented the idea of standardizing `unified executors', which was later abandoned~\cite{p0443}. In HPX, we have used the ideas outlined in that paper to develop the extensive customization mechanisms that we describe below.

All of the parallel algorithms in HPX rely on internally invoking a number of customization points that can be overloaded by either the executor or the execution parameters object (see Section~\ref{subsec:executors}). We call a \textbf{customization point} a function that can be overloaded by an external user-defined function or object such that it is selected at compile time as a replacement for an internal, predefined default implementation. Customization point objects are usually function object instances that fulfill the two objectives of a) unconditionally trigger (conceptified) type requirements on the arguments of that function, and b) dispatch to the correct function via argument dependent lookup (ADL). In C\texttt{++}, these are often dependent on a library feature that adds concept checking, resulting in, e.g. clearer compilation error messages in case of erroneous template instantiations. We describe the set of customization points used by HPX's standard algorithm implementation relevant to this work in Section~\ref{subsec:cpos}. In HPX, we rely on the \texttt{tag\_invoke} methodology~\cite{p1895} for the implementation of our customization points.

\subsubsection{HPX Executors and Execution Parameter Objects}
\label{subsec:executors}
In a previous paper~\cite{10.1007/978-3-031-97196-9_6}, we briefly explained HPX and Customization points. In this paper, we will elaborate on these topics and provide more details about executors and parameter objects which are crucial to our implementation. 

\textbf{Executors} provide a uniform interface for work creation and task scheduling by abstracting underlying resources where work physically executes. Underlying resources could be, for instance, a thread pool, SIMD units, GPU runtimes, or simply the current thread. In general, we call such resources \textbf{execution contexts}. As lightweight handles, executors impose uniform access to execution contexts. The uniformity of the API exposed by executors enables control over where work executes, even when it is executed and scheduled indirectly behind library interfaces. All of HPX's APIs related to executing threads, in particular all parallel algorithms, rely on executors for their thread scheduling needs. In a nutshell, executors are used to describe when, where, and how to execute a computation.

One notable advantage of utilizing executors is the ability to maintain optimizations external to and independently of the algorithm implementations, empowering users to decide when and where to leverage these optimizations. Consequently, the algorithm implementations themselves remains untouched. 

In HPX, several executors have been implemented to control various functionalities, such as restricting execution to a subset of the available processing units (cores) or to encapsulate different thread-scheduling implementations. The focus of this paper centers on an executor responsible for managing an optimal number of cores and optimal chunk sizes used while parallelizing the execution of map-style and map-reduce-style parallel algorithms.

\textbf{Execution Parameter Objects} provide a uniform way of customizing additional execution parameters, such as chunk sizes, priorities, thread affinities, or the number of processing units to be used for running a parallel algorithm. In this paper, we present an implementation of a new executor parameters object that adaptively controls the chunk sizes and the number of cores used to execute a given parallel algorithm.

The HPX execution policies expose additional APIs that allow the programmer to attach an executor (\texttt{.on()}) and/or execution parameter objects (\texttt{.with()}) as shown in Listing~\ref{lst:execpolicies}. Both APIs create new execution policies that encapsulate the given executor and/or execution parameters object. This design decision enabled a seamless extension of the well-documented parallel algorithm interface while providing deep and fine-grain customization capabilities to the user. 

\begin{lstlisting}[language=C++,frame=single,label=lst:execpolicies,caption=Customization examples for HPX execution policies,escapechar=!]
// use default execution policy execution::par 
// relying on default executor
std::vector<double> d = { ... };
fill(!\textbf{execution::par}!, begin(d), end(d), 0.0);
// rebind par to an user-defined executor
// using .on()
user_defined_executor ex = ...;
fill(!\textbf{execution::par.on(ex)}!, ...);
// rebind par to an user-defined executor and
// a user-defined execution parameters object
// using .on() and .with()
user_defined_params param = ...
fill(!\textbf{execution::par.on(ex).with(param)}!, ...);
\end{lstlisting}

In this paper, we introduce an executor that optimizes the number of cores and chunks for various workloads. Users can easily integrate this executor into their parallel algorithms using \texttt{.with()}. 


Our experiments focus on the parallel implementation of the \texttt{adjacent\_difference} algorithm in HPX. Listing~\ref{lst:example} demonstrates how users can attach this executor, referred to as \texttt{acc}, to the algorithm. 
Note that we pass \texttt{acc} explicitly by reference to support the use case described in Section~\ref{sec:motivation}. The first invocation of the algorithm will measure the time per loop execution and store the result in \texttt{acc}, thus making the time available to all subsequent invocations of the algorithm.

\begin{lstlisting}[language=C++,frame=single, label=lst:example,caption=Integrating the new execution parameters object (acc) with the adjacent-difference algorithm in HPX]
adaptive_core_chunk_size acc;
adjacent_difference(
    execution::par.with(std::ref(acc))),
    arr.begin(), arr.end(), res1.begin());
\end{lstlisting}


\subsubsection{Customization Points for HPX's Parallel Algorithms}
\label{subsec:cpos}

As outlined in Section~\ref{sec:motivation}, in order to optimize the execution times for parallel algorithms, two parameters need to be adaptively controlled: the size of the chunks, i.e. the size of the parts of the data array the algorithm is invoked on that are passed as separate concurrent tasks to a processing unit, and the number of processing units to utilize for a particular execution of a parallel algorithm. In this work we assume that the amount of work in the user-supplied loop body can be measured during the first invocation of a particular, repeatedly invoked parallel construct.

HPX's parallel algorithm implementations rely on invoking several customization points. We will focus, in particular, on three of those: \texttt{measure\_iteration}, \texttt{processing\_units\_count}, and \texttt{get\_chunk\_size} (see~\cite{kaiser_2024_598202}). Listing~\ref{lst:cpos} shows a simplified call sequence taken from the algorithm implementation of HPX. All of the customization points are called by passing the execution parameters object \texttt{params} and the executor \texttt{exec} that were bound to the execution policy used for invoking the parallel algorithm (see Section~\ref{subsec:algorithms}). This allows the programmer to provide the necessary execution context to the customization point implementations. The variable \texttt{loop\_body} is a function representing the user supplied loop body, while \texttt{count} holds the overall number of iterations.

\begin{lstlisting}[language=C++,frame=single, label=lst:cpos,caption=Exemplar invocation sequence of customization points in the implementation of HPX's parallel algorithms]
// Return the time per iteration for given
// loop body and overall number of iterations  
auto iteration_duration = measure_iteration(
    params, exec, loop_body, count);
// Return number of cores to utilize for
// given time per iteration and overall
// number of iterations
size_t cores = processing_units_count(
    params, exec, iteration_duration, count);
// Return size of chunks (i.e. number of
// iterations per chunk) to run as concurrent 
// tasks for given time per iteration, number
// of processing units, and overall number of
// iterations
size_t chunk_size = get_chunk_size(params,
    exec, iteration_duration, cores, count);
\end{lstlisting}

The expected semantics of the customization points are: 

\begin{itemize}
    \item \texttt{measure\_iteration} should return the time required to execute one iteration of the loop in some unspecified way. 
    \item \texttt{processing\_units\_count} should return the number of processing units the algorithm implementation should utilize while running the overall algorithm.
    \item  \texttt{get\_chunk\_size} should return the sizes of the array parts (chunks) that are run as concurrent tasks. 
\end{itemize}

The default implementations for these customization points perform minimal work resulting in splitting the array into chunks such that each of the available processing units receives an equal amount of iterations.

ere we present the results of implementing an execution parameter type \texttt{adaptive\_core\_chunk\_size} that exposes the aforementioned customization points by encapsulating the mathematical model we describe in Section~\ref{sec:theory}.

  \section{Experiments and Methodology}
Our new executor adaptively finds the optimal number of cores and the chunk size, so we call it \texttt{adaptive\_core\_chunk\_size} or (acc) in this paper.

In this execution parameters object, we have implemented the customization point---\texttt{measure\_iteration}---designed to compute the time per loop execution (See Listing \ref{lst:cpos}). The value of $T_1$ will be calculated once for each workload, the first time a loop is invoked. (Note that because the number of elements, $N_E$, may vary with different invocations of the loop, what we actually store is $T_E = T_1/N_E$. In this way, if future invocations of the loop have a different $N_E$, we can easily calculate the appropriate $T_1$ for that invocation.)

$T_0$ is an application constant value and it should be measured on a particular application and particular architecture. When we run an application on a particular architecture, we measure $T_0$ on the first invocation loop and use this value to optimize future invocations. We make sure this value is measured each time for each application separately since this value can be different for different loops.

The \texttt{processing\_units\_count} customization point (listing~\ref{lst:cpos}) uses $T_1$ and $T_0$ and applies Equation~\ref{eq:n} to find the optimal number of cores, $N$, for each workload. It then uses that value, unless it is more than the maximum available cores in the system, in which case the maximum available number of cores are used.
The last critical customization point implemented by this executor is \texttt{get\_chunk\_size}, which calculates the optimal chunk size based on the calculated number of cores (as previously returned from \texttt{processing\_units\_count}). This Customization point uses the minimum time per chunk calculated based on Eq~\ref{eq:t_m} and makes sure to return a chunk size ensuring the optimal result (see Section~\ref{subsec:cpos}).

\vspace{1em}

 For those more familiar with OpenMP than with HPX, the above implementation would provide an automatic way to pick the best value for the \textit{num\_threads} argument to the parallel pragma. Not only will this avoid slowdowns when loops are too small to benefit from parallelism, but it leaves cores available for other parallel tasks should they be needed.


In order to better understand the behavior of map-type algorithms, we conducted a thorough study of the effects of core count and chunk size on the \textit{adjacent-difference} algorithm, which is memory-bound, and an \textit{artificial-work} algorithm that is compute bound and an algorithm in ChplX.

Initially, we focused on executing the algorithm using executors with different numbers of cores and chunks. The experiments illuminate how different combinations of cores and chunks influence the algorithm's performance. In the next section, Experiment 1 (see Section~\ref{subsec:experiment1}), we examined the different conditions requiring varying numbers of chunks and cores for a set of workloads. We examined each characteristic (core and chunk size) separately to obtain the best combination. It became evident that for smaller input sizes, optimal performance is achieved with fewer cores and chunks. Conversely, employing more cores and chunks yields better performance for larger input sizes.

The rationale behind this is that for a smaller workload, an increased number of cores and chunks escalates overhead, hindering performance, while for larger workloads with a larger value of $N$, more chunks enables greater load balancing. To maintain optimal performance, we use an executor that automatically adjusts the optimal number of cores and chunks. 

\paragraph{\textbf{Experiment 1}}
\label{subsec:experiment1}
We evaluated all the performance measurements in this experiment using HPX V1.10.0 on a test machine with Intel Xeon Skylake processors, with 40 cores at 2.4GHz and 96 Gb of main memory, 2 sockets with 20 cores each, with hyperthreading disabled. In these experiments, each processor unit is the same as one core or thread. We used a benchmark to generate data and the result is the average of 50 iterations.

In the initial experiment, we kept the number of cores constant while varying the chunk sizes for various workloads. HPX has a very light-weight parallelism with very efficient work stealing~\cite{paper5364511}.  As a result, we expect to see good resource utilization. However, caution is needed, as excessive chunking can introduce significant overhead. The graphs also indicate the best chunking options for each number of cores used.

\begin{figure}[tpb]
    \centering
    \begin{subfigure}[b]{0.48\textwidth} 
        \centering
        \includegraphics[width=\linewidth,trim=0 0 20 0]{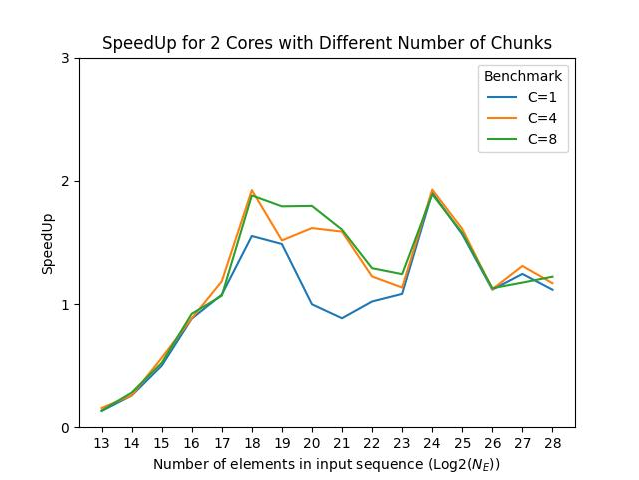}
        \caption{\footnotesize Speedup with 2 processing units (cores)}
        \label{fig:chunking1}
    \end{subfigure}
    \hfill
    \begin{subfigure}[b]{0.48\textwidth} 
        \centering
        \includegraphics[width=\linewidth,trim=0 0 20 0]{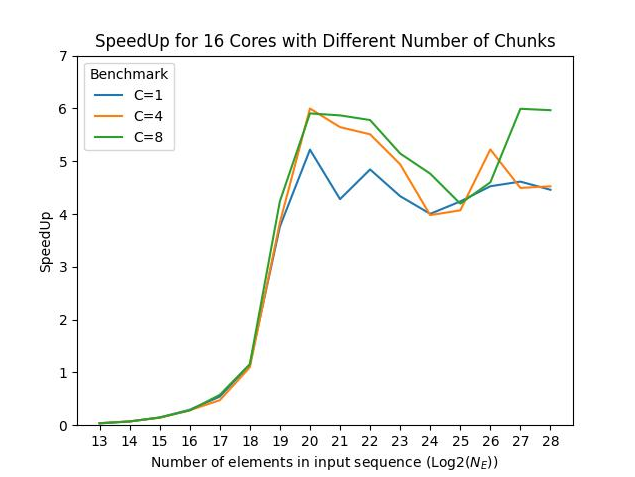}
        \caption{\footnotesize Speedup with 16 processing units (cores)}
        \label{fig:chunking2}
    \end{subfigure}
    
    \vspace{0.5cm} 

    \begin{subfigure}[b]{0.48\textwidth} 
        \centering
        \includegraphics[width=\linewidth,trim=0 0 20 0]{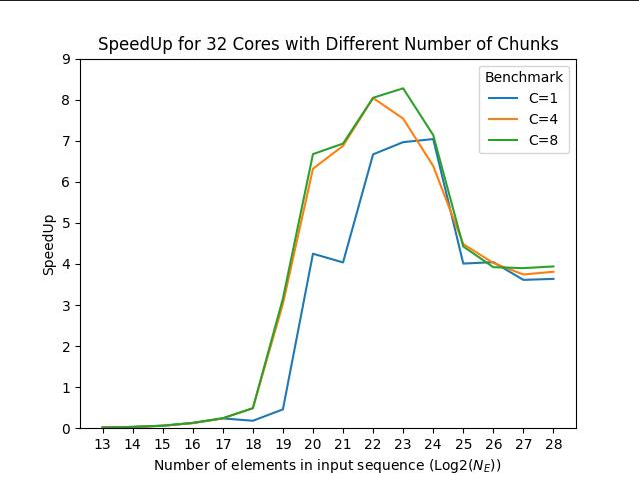}
        \caption{\footnotesize Speedup with 32 processing units (cores)}
        \label{fig:chunking3}
    \end{subfigure}
    \hfill
     \begin{subfigure}[b]{0.48\textwidth} 
        \centering
        \includegraphics[width=\linewidth,trim=0 0 20 0]{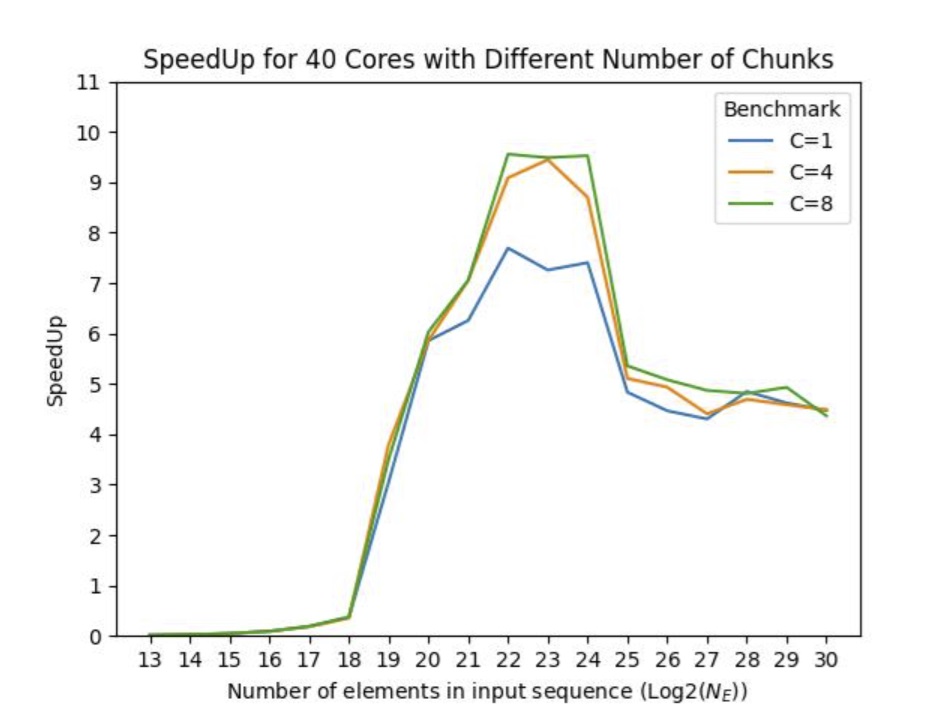}
        \caption{\footnotesize Speedup with 40 processing units (cores) (New)}
        \label{fig:chunking4}
    \end{subfigure}
      
    \vspace{0.5cm} 

    \begin{subfigure}[b]{0.6\textwidth} 
        \centering
        \includegraphics[width=\linewidth,trim=0 0 20 15]{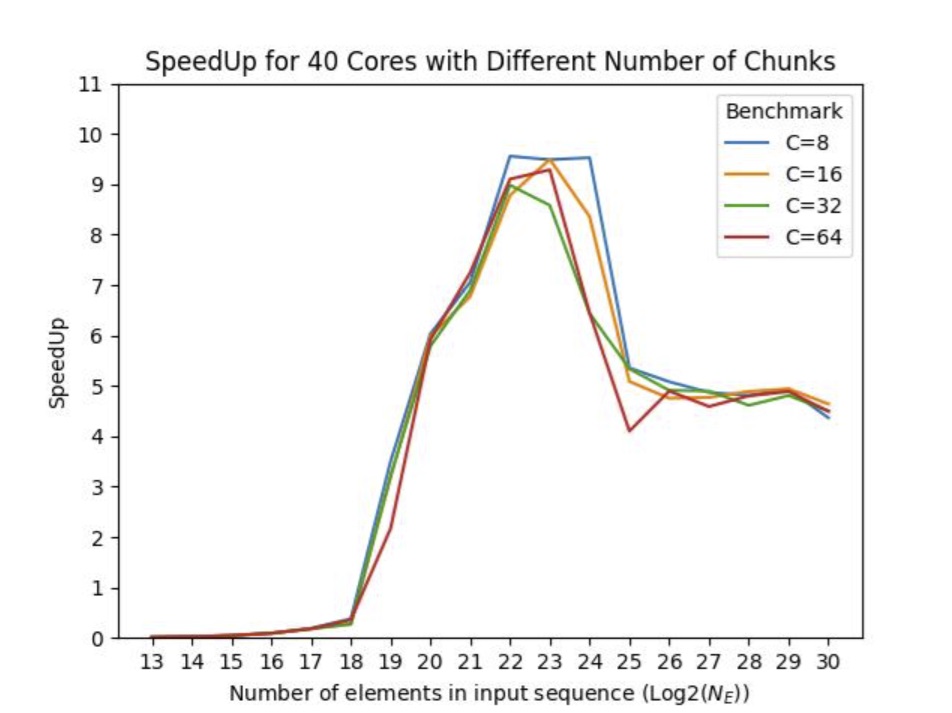}
        \caption{\footnotesize Speedup with 40 processing units (cores) (New)}
        \label{fig:chunking5}
    \end{subfigure}

    \caption{\footnotesize Array size vs. speedup when using different numbers of processing units (cores) for parallelizing the finite-difference algorithm for different numbers of chunks-per-core, $C$. 
    For comparison, the value of $C$ in these runs behaves like the chunk size argument to OpenMP's static scheduling algorithm.} 
    \label{fig:chunking}
\end{figure}

In our first paper~\cite{10.1007/978-3-031-97196-9_6}, we evaluated core counts of 2, 16, and 32 with three possible chunking configurations: a number of chunks equal to the number of cores ($C=1$), four times the number of cores ($C=4$), and eight times the number of cores ($C=8$). See Figure~\ref{fig:chunking1}, \ref{fig:chunking2}, and~\ref{fig:chunking3}.
In this paper, we extended our evaluation to utilize all available cores on the test machine (see Figure~\ref{fig:chunking4}). The results indicate that, among the fixed chunk configurations, $C=8$ consistently yielded the best overall performance.
We also explored finer-grained chunking by comparing the optimal setting of $C=8$ against larger values---specifically, 16, 32, and 64 times the number of cores ($C=16,32,64$)---as shown in Figure~\ref{fig:chunking5}.

\begin{figure}[tpb] 
  \centering 
  \includegraphics[width=0.7\linewidth,trim=0 0 20 0]{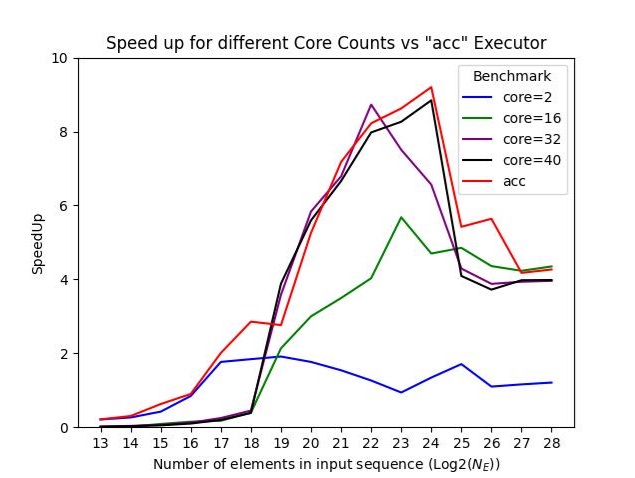} 
  \caption{\footnotesize Speedup measured for the adjacent difference algorithm across a range of core counts and input sizes. We compare executions (for different numbers of cores) with the results measured when using the new \texttt{adaptive\_core\_chunk\_size} \textbf{(acc)} ({\color{red}red line}). }
  \label{fig:chunking6}
\end{figure}

\paragraph{\textbf{Experiment 2}}
In the second experiment, we compared the different fixed core counts using $C=4$ (current HPX implementation) along with the result with our adaptive algorithm using adapted cores and $C=8$ which was the most optimal chunking (\textit{acc}). See Figure~\ref{fig:chunking6}. This is equivalent to using OpenMP and providing static scheduling and a \textit{num\_threads} argument. All the performance measurements in this experiment are the same as Experiment 1. 

The results show that using all available cores for every workload is not always the best option. The graph indicates that, for smaller workloads, using fewer cores is more effective, while larger workloads benefit from using more cores.

\paragraph{\textbf{Experiment 3}}
In our initial result~\cite{10.1007/978-3-031-97196-9_6}, we ran this experiment on two different machines. In this paper we use three different types of hardware: one is the same as experiment one (Intel Hardware); the second is AMD EPYC processors with 48 cores, 2 sockets with 24 cores each; and the third (the new one) is a RISC-V architecture with 64 cores and four Numa nodes. However, to evaluate the impact of our \texttt{adaptive\_core\_chunk\_size} executor on compute-bound loops, we used an artificial workload instead of \textit{adjacent-difference}. We then compared the performance of our new executor against the default parallel execution policies (used by both OpenMP and HPX) across varying workloads.

\begin{figure}[tpb]
    \centering
    \begin{minipage}{0.49\linewidth}
        \centering
        \includegraphics[width=\linewidth,trim=0 0 20 0]{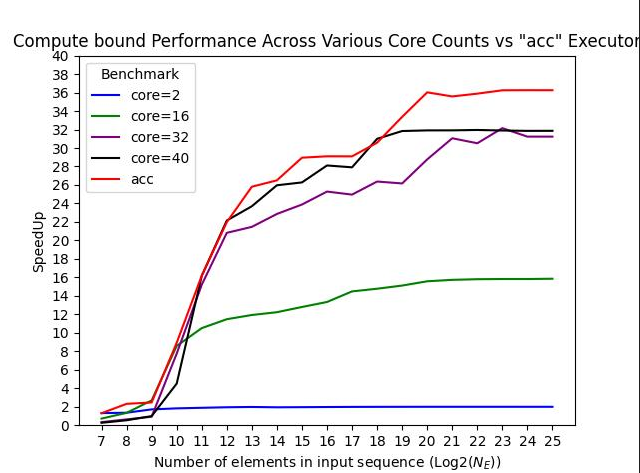}
        \caption{\footnotesize Speedup across various core counts for a \textbf{compute-bound} use case when using the default static parameters compared to using the new \texttt{adaptive\_core\_chunk\_size}\textbf{(acc)}({\color{red}red line}) across varying input sizes on Intel hardware.} 
        \label{fig:compute}
    \end{minipage}
    \hfill
    \begin{minipage}{0.49\linewidth}
        \centering
        \includegraphics[width=\linewidth,trim=0 0 20 0]{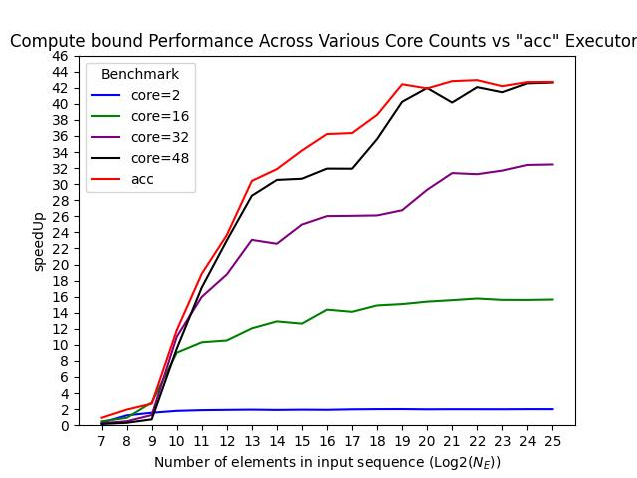}
        \caption{\footnotesize Speedup across various core counts for a \textbf{compute-bound} use case when using the default static parameters compared to using the new \texttt{adaptive\_core\_chunk\_size}\textbf{(acc)}({\color{red}red line}) across varying input sizes on AMD hardware.} 
        \label{fig:compute_buran}
    \end{minipage}
    
     \vspace{0.5cm} 

    \begin{minipage}{0.49\linewidth} 
        \centering
        \includegraphics[width=\linewidth,trim=0 0 20 15]{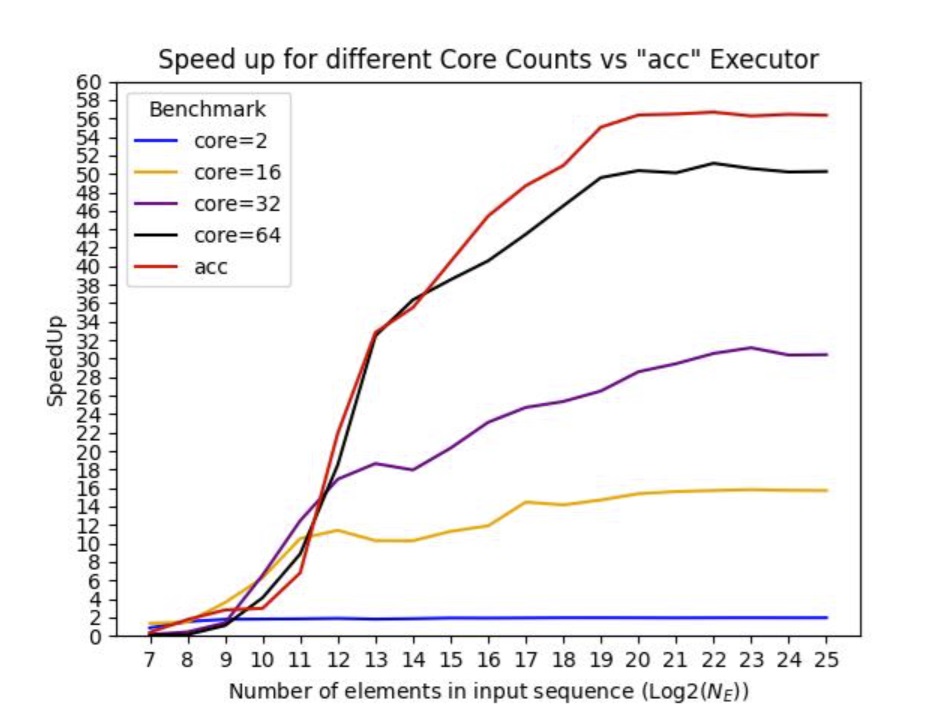}
        \caption{\footnotesize Speedup across various core counts for a \textbf{compute-bound} use case when using the default static parameters compared to using the new \texttt{adaptive\_core\_chunk\_size}\textbf{(acc)}({\color{red}red line}) across varying input sizes on RISC-V hardware. (New)} 
        \label{fig:risc5}
    \end{minipage}
\end{figure}

Figure~\ref{fig:compute}, \ref{fig:compute_buran} , and~\ref{fig:risc5} present this comparison. In these graphs, parallelization begins earlier and scales more rapidly. As before, $T_1$ represents the total execution time. However, this algorithm scales differently compared to the previous case (bigger $T_1$ for the same input size). As a result, the \texttt{adaptive\_core\_chunk\_size} execution parameters object not only starts parallelization on smaller arrays, but it also uses all available cores in the system earlier than in the default case. As a result, especially very small workloads show significant benefits from dynamic core adjustments with our new executor. Additionally, our executor consistently outperforms other core configurations in larger workloads. This performance gain is attributed to its ability to select optimal chunk sizes compared to the default parallel policies in HPX. 

Figure~\ref{fig:risc5} presents our new experiment conducted on a RISC‑V machine. We observed that the time per iteration on this platform is noticeably higher than on our other test machines, which may be attributed to specific architectural constraints of RISC‑V systems~\cite{diehl2023evaluating}. Regarding this characteristic---namely, that RISC-V is inferior in computation time compared to platforms like AMD~\cite{_Volokitin_2023}---our executor still outperforms other experimental cases.

\begin{figure}[tpb] 
  \centering 
  \includegraphics[width=0.7\linewidth,trim=0 0 20 0]{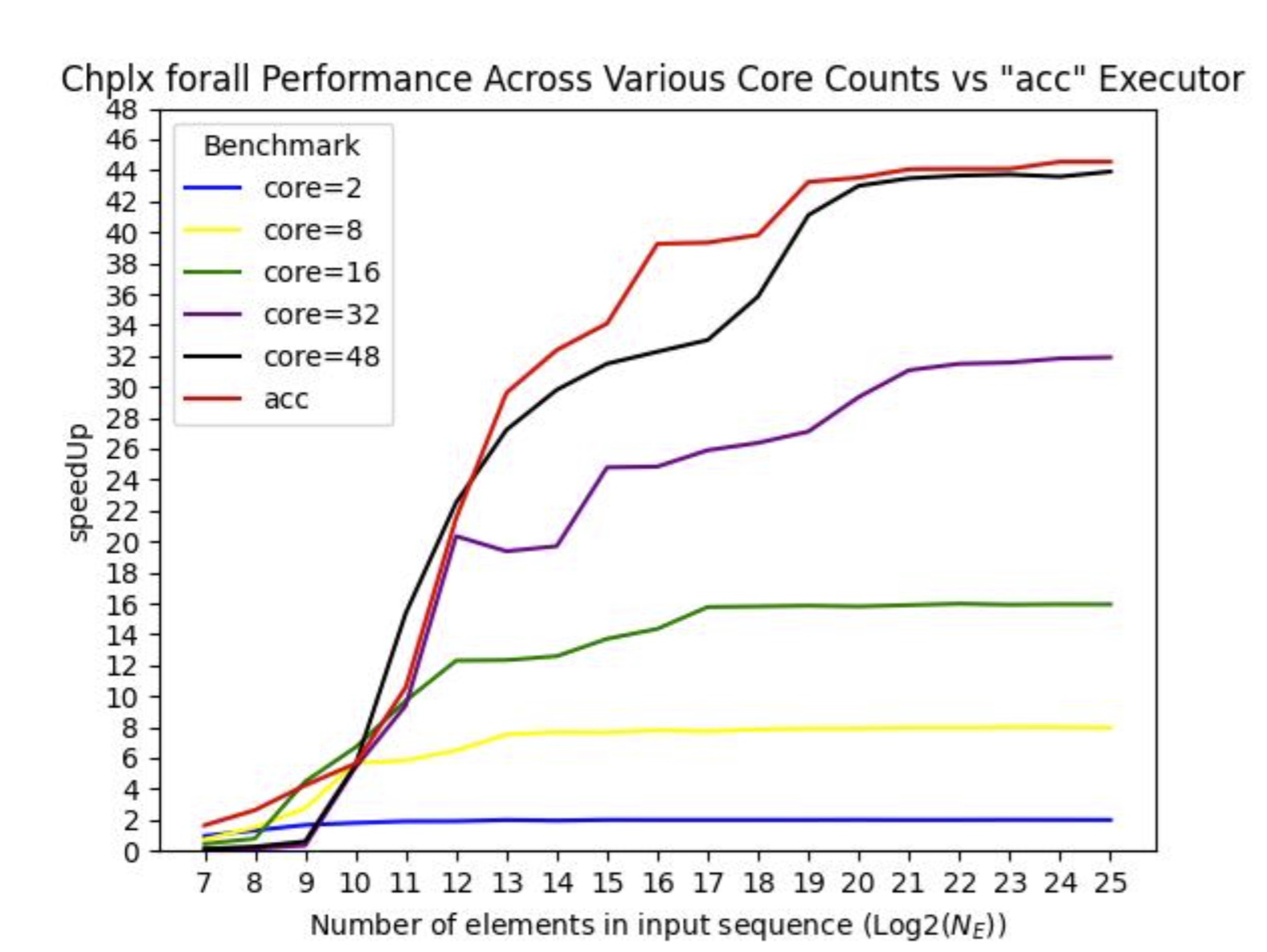} 
  \caption{\footnotesize Speedup measured for the Chplx \texttt{forall} across a range of core counts and input sizes. We compare executions (for different numbers of cores) with the results measured when using the new \texttt{adaptive\_core\_chunk\_size} \textbf{(acc)} ({\color{red}red line}). (New) }
  \label{fig:ChplX}
\end{figure}

\paragraph{\textbf{Experiment 4}}
In this new set of experiments, we used the ChplX library, which is an abstraction that mimics chapel's semantics~\cite{atre2025closingsourcecomplexitygap}. It is designed for a line by line source to source translation of Chapel programs to C\texttt{++} with HPX as its runtime system. More about this is explained in~\cite{atre2025closingsourcecomplexitygap} paper. In this library, we chose ChplX \texttt{forall} as an algorithm for our new experiment. Figure \ref{fig:ChplX} shows the comparison between our new executor \texttt{acc} with different number of cores and static number of chunks used in HPX. The machine used for this experiment is AMD as defined earlier.

  \section{Results}




  

 In this section, we present and analyze the findings from our experiments. Our goal was to identify the optimal core-chunk combinations for a map-style algorithm and develop an \texttt{adaptive\_core\_chunk\_size} execution parameters object.

\noindent
\textbf{Performance analysis of varying chunk sizes.}
In the first experiment, we varied chunk sizes while keeping the number of cores constant. The results, illustrated in Figures~\ref{fig:chunking1},~\ref{fig:chunking2},~\ref{fig:chunking3}, \ref{fig:chunking4}, and~\ref{fig:chunking5} highlight how performance varies with chunk sizes across various workloads. Setting the number of chunks to eight times of number of cores is always the better option. 


\begin{figure}[tpb]
    \centering
    \begin{subfigure}[b]{0.48\textwidth} 
        \centering
        \includegraphics[width=\linewidth,trim=0 0 20 0]{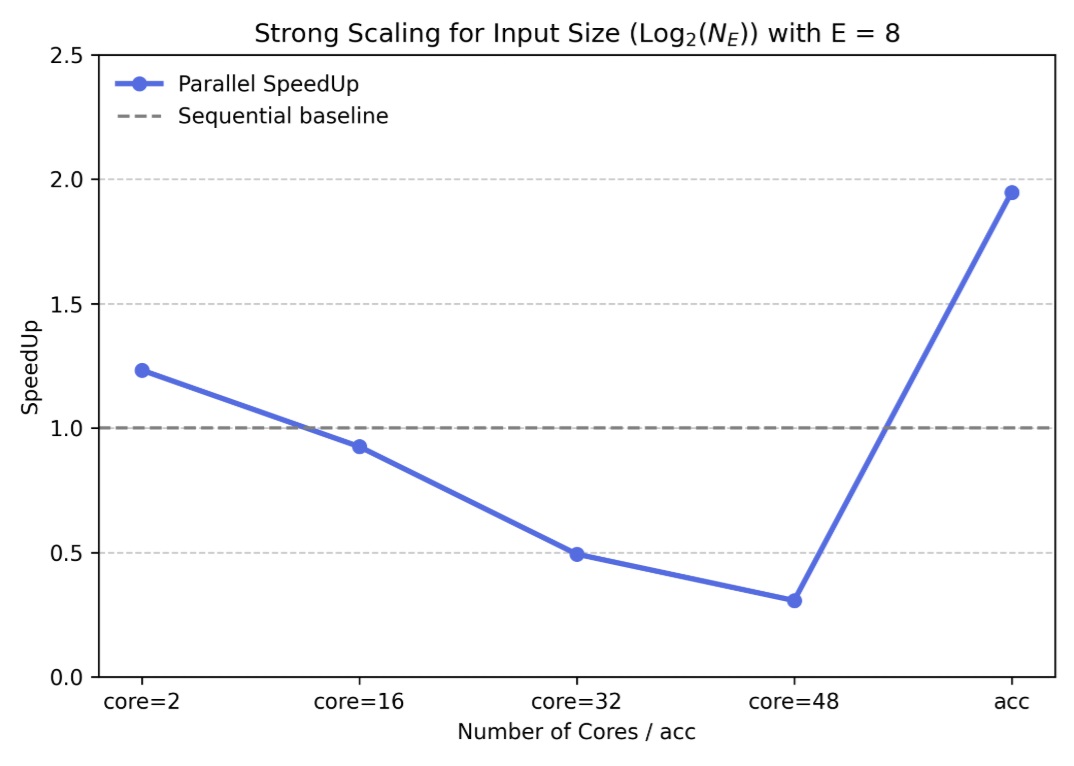}
        \caption{\footnotesize Strong Scaling for different core counts vs \texttt{acc} Executor}
        \label{fig:strong1}
    \end{subfigure}
    \hfill
    \begin{subfigure}[b]{0.48\textwidth} 
        \centering
        \includegraphics[width=\linewidth,trim=0 0 20 0]{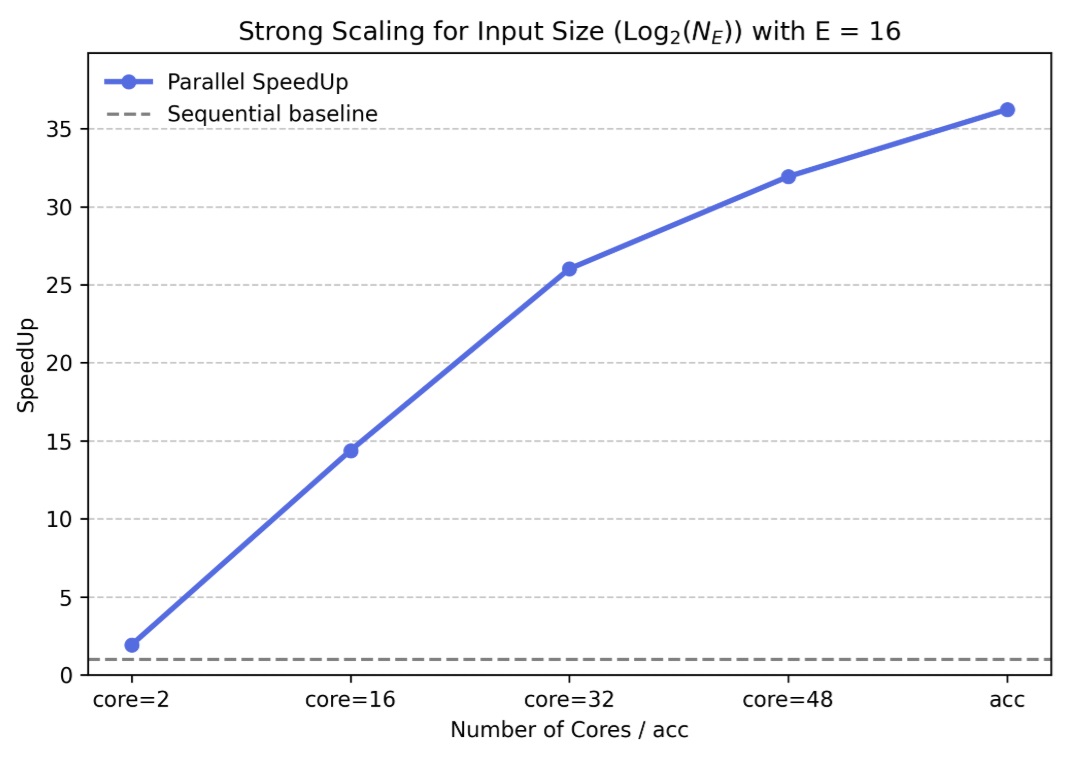}
        \caption{\footnotesize Strong Scaling for different core counts vs \texttt{acc} Executor}
        \label{fig:strong2}
    \end{subfigure}
    
    \vspace{0.5cm} 

    \begin{subfigure}[b]{0.6\textwidth} 
        \centering
        \includegraphics[width=\linewidth,trim=0 0 20 15]{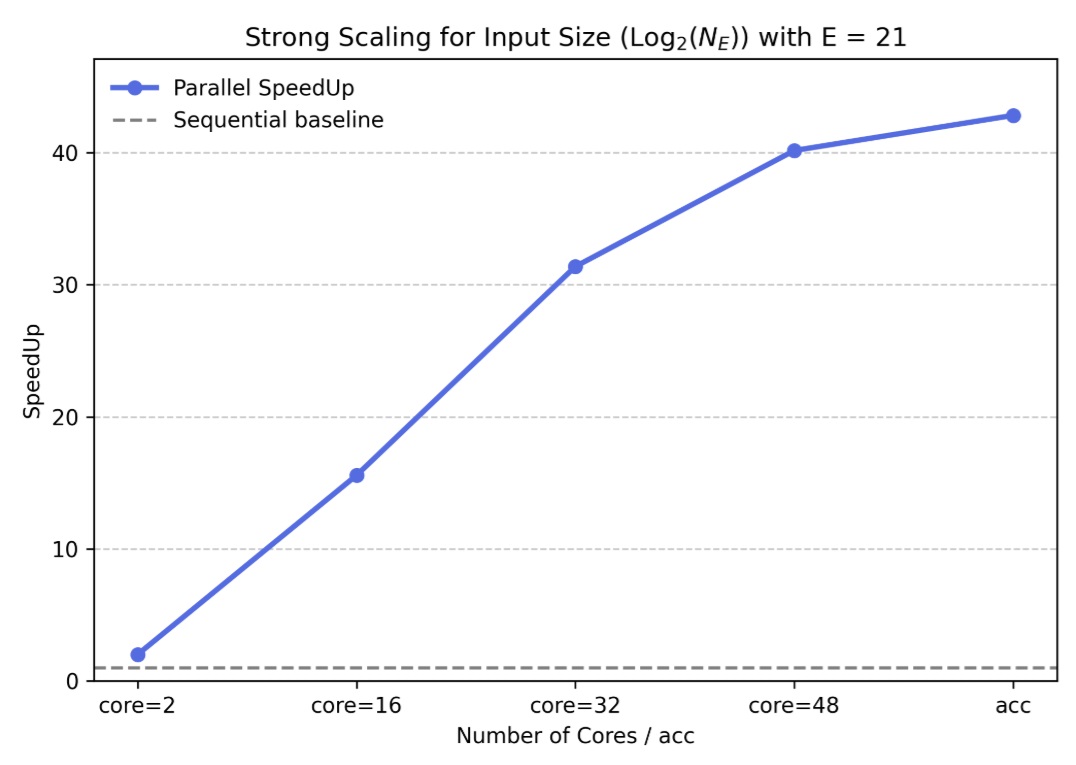}
        \caption{\footnotesize Strong Scaling for different core counts vs \texttt{acc} Executor)}
        \label{fig:strong3}
    \end{subfigure}

    \caption{\footnotesize Strong scaling results for three input sizes.
Speedup is plotted against core count (including \texttt{acc}) for fixed input sizes labeled in $log_2$ scale. The sequential baseline is normalized to 1. (New)}
    \label{fig:strong}
\end{figure}

\noindent
\textbf{Performance of adaptive chunking.}
Building on the insights from the two experiment sets, we developed a
customization point object \texttt{adaptive\_core\_chunk\_size}. This object dynamically adjusts core and chunk configurations based on workload size. Figure~\ref{fig:chunking6} illustrates the performance of various processor counts alongside the results of our executor. The red line (acc) optimally selects the number of cores and chunks for each workload. By dynamically adjusting the number of processors---utilizing fewer for smaller workloads and more for larger ones---this approach demonstrates improved overall performance across all workloads. 

 \noindent It is also worth noting that in Experiment 2, Figure~\ref{fig:chunking6}, although our new executor achieves the best overall performance, the parallel speedup remains limited, even with 40 cores. Specifically, we observe only approximately a 10x speedup compared to the sequential execution. This is primarily because the algorithm \texttt{adjacent\_difference} is a memory-bound algorithm and there is not enough work to maintain 95\% efficiency. However, when the algorithm is changed to a compute-bound one, as shown in Figures~\ref{fig:compute}, \ref{fig:compute_buran}, and~\ref{fig:risc5} the benefits of parallelism become more evident. On a 40-core machine, we achieve up to 38x speedup, on a 48-core machine, the speedup reaches up to 46x, and on a 64-core machine, we achieve up to 57x speedup compared to the sequential execution.

 In our new experiments, We also used strong scaling as a tool to illustrate how core count affects parallel speedup and to compare the performance of our new executor, \texttt{acc}, against standard configurations. See Figure~\ref{fig:strong}. We selected three input sizes---ranging from small to large---to represent different workload scales. The strong scaling results demonstrate that for smaller input sizes, using a large number of cores is suboptimal, whereas larger inputs benefit from increased core counts. Our adaptive executor, \texttt{acc}, consistently achieved the best speedup by adjusting the core count accordingly: it selected fewer cores for smaller workloads and more cores for larger ones. This adaptive behavior highlights \texttt{acc}'s ability to match resource usage to workload characteristics, maximizing performance across varying scenarios.

\noindent In summary, the new approach is beneficial in three key ways. Firstly, through \textbf{performance gain,} the adaptive executor consistently outperformed static configurations. By dynamically tuning core and chunk sizes, it minimized overheads and maximized resource utilization, leading to significant performance improvements across all workload sizes. Secondly, through \textbf{overhead reduction,} for small workloads, the adaptive executor effectively reduced overhead by limiting cores and chunks. For large workloads, it leveraged maximum available cores and increased chunk count, enhancing parallelism. Thirdly, \textbf{through daptability,} the adaptive executor's ability to adjust to varying workloads makes it highly versatile, suitable for a wide range of parallel applications.

\section{Conclusions}
The development and optimization of parallel algorithms are crucial for maximizing computational resource efficiency. This paper studies map-type algorithms and develops a mathematical model to improve their performance based on runtime parameters. The focus of the study is on determining the optimal combination of core counts and chunk sizes at runtime for various workloads. The model has been implemented and integrated into the HPX C++ library. By introducing the HPX execution parameters object \texttt{adaptive\_core\_chunk\_size}, we have made it possible to dynamically adjust the number of cores and the size of the chunks based on the application's workload size and the available number of cores in the execution environment. The result is improved performance and scalability across a wide range of workloads for parallel algorithms. This adaptive approach reduces the overheads and enhances resource utilization, leading to better overall performance. This is particularly true for smaller workloads. 
 \section{Future Work}
 
Future work will focus on several areas to build upon the findings of this paper. First, we aim to explore the application of the \texttt{adaptive\_core\_chunk\_size} execution parameters object to more algorithms. This includes investigating its impact on additional memory-bound and I/O-bound algorithms to assess its versatility and performance benefits across different computational paradigms. Second, fluctuations in the performance graphs suggest the influence of cache effects. Future models can factor in these effects by leveraging measurement tools like PAPI.
In this work, we develop an abstract model optimizing the performance of parallel loop construct in general cases. We also showed the models applicability for a set of artificial use cases. Applying this model to a real world application is part of our planned future work and is thus beyond the scope of this paper. We plan to conduct comprehensive benchmarking on a variety of real-world applications to validate the executor's effectiveness and robustness. This includes performance evaluations on different hardware architectures, such as GPUs and specialized accelerators, ensuring the executor's adaptability and efficiency across diverse computing environments. By testing on a broader range of platforms, we aim to demonstrate the generalizability and scalability of our approach. We also plan to integrate machine learning techniques to enhance the executor's decision-making process. By training models on various workloads and system configurations, the executor can predict and adapt to optimal core and chunk sizes more accurately and dynamically. This approach could further reduce overheads and improve performance, particularly in heterogeneous computing environments.
Moreover, additional experiments across different architectures, algorithms, and libraries revealed that while \texttt{acc} consistently performs well over a range of input sizes, there are still cases where it could make more optimal decisions regarding core selection. We plan to further investigate and refine the underlying formula to improve its performance across all input sizes.

  \bibliographystyle{unsrt}
\bibliography{refs.bib}
\end{document}